\documentclass[a4paper,fleqn,usenatbib]{mnras}

\usepackage[T1]{fontenc}
\usepackage{ae,aecompl}

\usepackage{graphicx}	% Including figure files
\usepackage{amsmath}	% Advanced maths commands
\usepackage{amssymb}	% Extra maths symbols
\usepackage{adjustbox}
\usepackage{multicol}        % Multi-column entries in tables
\usepackage{bm}		% Bold maths symbols, including upright Greek

\title[Differential rotation in K, G, F and  A stars]
{Differential rotation in K, G, F and A stars}
\author[L.A. Balona, O. P. Abedigamba] 	
{L. A. Balona$^1$, O. P. Abedigamba$^2$\\
\\
$^1$South African Astronomical Observatory, P.O. Box 9, Observatory, Cape
Town, South Africa\\
$^2$Department of Physics, North-West University, Private Bag X2046, Mmabatho
2735, South Africa
}

\begin{document}

\date{Accepted .... Received ...}

\pagerange{\pageref{firstpage}--\pageref{lastpage}} \pubyear{2011}

\maketitle

\label{firstpage}

\begin{abstract}
Rotational light modulation in {\it Kepler} photometry of K -- A stars is used
to estimate the absolute rotational shear.  The rotation frequency spread in 
2562 carefully selected stars with known rotation periods is measured using 
time-frequency diagrams.  The variation of rotational shear as a function of 
effective temperature in restricted ranges of rotation period is determined.  
The shear increases to a maximum in F stars, but decreases somewhat in the A
stars.  Theoretical models reproduce the temperature variation quite well.  
The dependence of rotation shear on rotation rate in restricted temperature 
ranges is also determined.   The dependence of the shear on the rotation rate 
is weak in K and G stars, increases rapidly for F stars and is strongest in A 
stars.  For stars earlier than type K, a discrepancy exists between the 
predicted and observed variation of shear with rotation rate.  There is a 
strong increase in the fraction of stars with zero frequency spread with 
increasing effective temperature.  The time-frequency diagrams for A stars are
no different from those in cool stars, further supporting the presence of 
spots in stars with radiative envelopes.
\end{abstract}


\section{Introduction}

The rotation period of an individual sunspot depends on it latitude: the
higher the latitude, the longer the period.  Examination of spectral line
Doppler shifts at the limb of the Sun shows that the rotational velocity
has the same latitude dependence as sunspot rotation rates.  In addition, 
helioseismic measurements have revealed the internal rotation profile of the 
Sun.  This shows that differential rotation varies with depth as well as
latitude. However, the radiative inner core of the Sun appears to rotate 
almost as a solid body \citep{Thompson2003}.

Most cool stars, either fully convective or with a convective envelope like 
the Sun, will have spots on their surfaces, giving rise to light variations.
The rotation periods can be obtained from a time series of photometric
observations.  Over 500 stars of spectral types F--M observed from the ground 
are classified as rotational variables \citep{Strassmeier2009}.  The 
{\it Kepler} mission has been a  very fruitful source for the study of 
starspots and the rotation periods of many thousands of cool {\it Kepler} 
stars have been measured \citep{McQuillan2013, McQuillan2014, Reinhold2013, 
Nielsen2013}.

Surface differential rotation can be inferred using three different 
techniques.  The first technique is Doppler imaging (DI). In DI the location 
of individual spots can be estimated from their effect on the spectral line 
profiles, provided the star is rotating at a sufficiently high rate 
\citep{Collier2002}.  A time series of Doppler images obtained with high 
dispersion and high signal to noise (S/N) allows differential rotation to be 
measured from differences in the rotation periods of individual spots at 
different latitudes. The second technique is the Fourier transform (FT)
method \citep{Reiners2002}.  In the FT method, the Doppler shift at different 
latitudes due to rotation can be inferred from the Fourier transform of
the line profiles.  Note that the star does not need have starspots for this
method to be used, but a spectrum with very high S/N is required.
The method is limited to stars with moderate and rapid rotation.  The big 
advantage is that latitudinal differential rotation can be measured from a 
single exposure.  The third method, time series photometry, has already been 
mentioned.  By following the variation of rotation period over time, a lower 
limit can be estimated for the rotational shear.  If there is more than one
starspot at different latitudes, photometric measurements will show close 
multiple periods. The range in periods provides a lower limit to the
rotational shear \citep{Reinhold2013}.

The latitudinal differential rotation in stars is usually described by a 
law of the form $\Omega(\theta) = \Omega_e (1 - \alpha\sin^2\theta)$, where
$\theta$ is the latitude and $\Omega_e$ the angular rotation rate at the 
equator. The rotational shear is the difference in rotation rate between the 
equator and pole, $\Delta\Omega = \Omega_e - \Omega_p = \alpha\Omega_e$.
For the Sun $\alpha = 0.2$.  Differential rotation is categorized as 
``solar-like'' when $\alpha > 0$, or as ``antisolar'' when the polar regions 
rotate faster than the equator ($\alpha < 0$).  In order to determine 
$\alpha$, the latitude of a spot needs to be known.  This can only be done 
with the DI method.  Photometric methods can only provide a lower 
limit of $\Delta\Omega$.  It is difficult to determine the sign of $\alpha$ 
and usually only the absolute value can be inferred. 

Investigation of differential rotation in stars has a long history. 
\citet{Henry1995} monitored the light curves of four active binary stars and 
found that their rotation periods changed with time, suggesting differential 
rotation.  From photometric time series observations of 36 cool stars, 
\citet{Donahue1996} found that the range of period variation increases with 
increasing period.  \citet{Reiners2003} used the FT method to estimate 
$\alpha$ in 32 stars of spectral types F0--G0.  Non-zero values of $\alpha$ 
were obtained for 10 stars.  They find that differential rotation appears to 
be more common in slower rotators and that the rotational shear depends on the 
rotation period.

\citet{Barnes2005} applied the DI technique to 10 G2--M2 stars and found that 
rotational shear increases with effective temperature.  They find a weak 
dependence on rotation rate.  Using the FT method,  \citet{Reiners2006} 
concluded that A stars generally rotate as solid bodies.  Using the same
method, \citet{AmmlervonEiff2012} were able to measure differential rotation 
in 33 A--F stars. They found evidence for two populations of differential 
rotators: one of rapidly rotating late A stars at the granulation boundary 
with strong horizontal shear and the other of mid- to late-F type stars with 
moderate rates of rotation and less shear. 

The photometric method in which a periodogram is used to detect multiple
close rotation periods has been used by \citet{Reinhold2013} for {\it Kepler} 
stars with $T_{\rm eff} < 6000$\,K.  They find that $\alpha$ increases with 
rotational period, and slightly increases towards cooler stars. The absolute 
shear shows only a weak dependence on rotation period over a large period 
range. More recently, \citet{Reinhold2015} confirmed that $\alpha$ increases 
with rotation period for stars with $T_{\rm eff} < 6700$\,K, while hotter 
stars show the opposite behaviour.

In summary, there is general agreement that the rotational shear,
$\Delta\Omega$, is low in M and K stars and reaches a maximum in F stars.
The dependence of shear with period seems to be weak.  The DI and FT methods 
require expensive resources and cannot be applied to stars with slow rotation.
{\it Kepler} photometry provides an invaluable resource, consisting of nearly 
four years of almost uninterrupted photometric measurements with 
unprecedented precision at a cadence of 30\,min.  While progress has been 
made in investigating differential rotation in these stars 
\citep{Reinhold2013, Reinhold2015}, the results are somewhat confusing.

In this paper we selected a sample of {\it Kepler} stars with known rotation
periods, but excluded pulsating stars and eclipsing variables.  A different 
approach is used to estimate the rotational shear.  Instead of relying on a 
secondary period, we construct a time-frequency diagram for each star.  In 
this way, spots with variable frequencies and amplitudes can be recognized by 
visual inspection of the time-frequency diagram presented as a grayscale image.
The total spread in rotation frequency for each star is estimated, giving 
a lower bound for $\Delta\Omega$.  In addition, the analysis is extended to 
A stars which in the past have been neglected in the belief that starspots 
should not exist in stars with radiative envelopes.

\section{The data}

{\it Kepler} light curves are available as uncorrected simple aperture 
photometry (SAP) and with pre-search data conditioning (PDC) in which
instrumental effects are removed \citep{Stumpe2012, Smith2012}.  The vast 
majority of the stars are observed in long-cadence (LC) mode with exposure 
times of about 30\,min.  These data are publicly available on the Barbara A. 
Mikulski Archive for Space Telescopes (MAST, {\tt archive.stsci.edu}).  
We used all available PDC data for each star.

The sample is taken from a list of over 20000 {\it Kepler} stars which 
consists of all stars observed in long-cadence mode (30\,min cadence) and 
$T_{\rm eff} > 6500$\,K and nearly all stars with {\it Kepler} magnitude 
$K_p < 12.5$\,mag.  We classified these stars according to variability type.
Most of the stars with $6500 < T_{\rm eff} < 7500$\,K could not be used
because they have multiple stable frequencies and amplitudes. These are the 
pulsating $\gamma$~Doradus variables.  The $\delta$~Scuti stars, which are the 
only other type of pulsating main sequence star later than type B, are easily 
identified by the presence of high frequencies.  These were excluded as were
eclipsing binaries.

Of the remaining stars, only those appearing in the catalogues of rotation
periods by \citep{McQuillan2014} or \citet{Reinhold2013} (stars cooler than 
7500\,K) or \citet{Balona2013c} (A-type stars) were selected.  The stars are
all on the main sequence with rotation periods shorter than about 30\,d. 

\section{Method}

 Because we expect starspots to migrate in latitude, the rotation 
frequency may slightly change with time.  This results in a broadening of the 
rotational peak in the periodogram of the whole data set.  If the spot
changes size or contrast, the resulting change in light amplitude will also
cause line broadening.  These effects tend to reduce the visibility of a
spot in the periodogram and the estimated frequency spread will be reduced.
To avoid this problem, we need to sample the frequency and amplitude of the 
spot over a sufficiently small time interval.  The evolution of frequency and 
amplitude with time can be constructed by sampling the periodogram over a 
limited time span at regular intervals over the whole light curve.  The time 
span used for the periodogram is called the ``window size''. 

The choice of window size requires some consideration.  If it is small, time 
resolution is good, but frequency resolution is poor and vice-versa.  After 
some trial and error, we settled on a window size of 200\,d and a step size of 
20\,d. In calculating the time-frequency diagram we used a weighting scheme in
which points further from a given time step are given lower weights in 
accordance with a Gaussian.  We chose the FWHM of the Gaussian to be half the 
window size (i.e. FWHM = 100\,d).  However, the result does not depend very 
much on the FWHM or even if the points are given the same weight.  The window 
size of the first and last sampling times are, of necessity, only half the 
chosen window size.  It is possible to use the full window size only after 
the first 100\,d and before the last 100\,d in the data set.  The Lomb-Scargle
periodogram for unequally-spaced data \citep{Press1989} was used in
constructing the time-frequency diagram. 

The time-frequency diagram can be displayed as a grayscale image in which
a sinusoidal signal of constant period and amplitude appears as a sharp line 
of constant intensity.  A spot which drifts in latitude and changes in size 
or contrast will appear as a curve with variable intensity.  The 
time-frequency diagram thus displays how the spot evolves in period and 
amplitude and can be used to discriminate between a starspot and a
pulsational signal, for example.  Furthermore, spots which might exist for 
a short period, and which would not necessarily give rise to a significant
secondary peak in the periodogram, will be easier to detect.  

\begin{figure}
\centering
\includegraphics[]{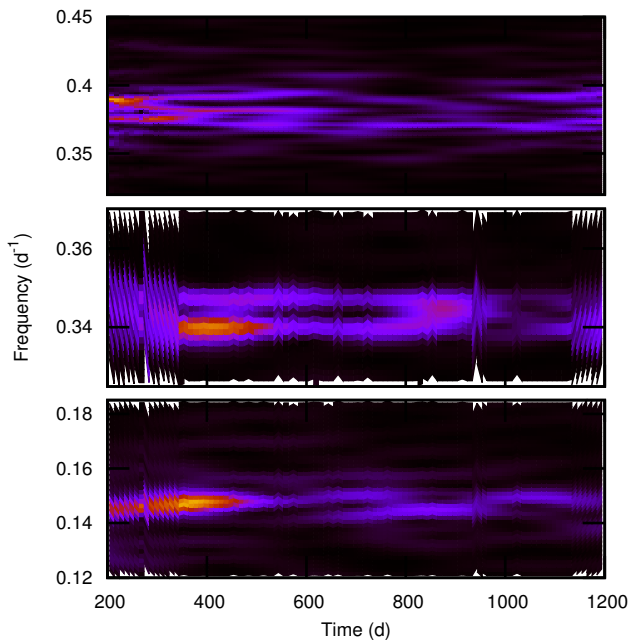}
\caption{Examples of time-frequency grayscale plots.  The top panel is for
KIC\,7135294 ($T_{\rm eff} = 5953$\,K), the middle panel KIC\,12061741 
($T_{\rm eff} = 9169$\,K) and the bottom panel KIC\,3122749 ($T_{\rm eff} =
9745$\,K).}
\label{gray}
\end{figure}

Examples of time-frequency grayscale plots are shown in Fig.\,\ref{gray}. 
As the figure shows, the grayscale image does not show just a single line, but
a band of closely-spaced and often interweaving lines of variable intensity 
caused by spots drifting in latitude and of finite lifetime.  A lower limit of
the rotation shear, $\Delta\Omega$, is obtained from the frequency
difference between the highest and lowest frequencies.  The shear can be
measured in 2562 stars.  For most of these stars, the first harmonic of the 
rotation frequency is also visible.

\begin{figure}
\centering
\includegraphics[]{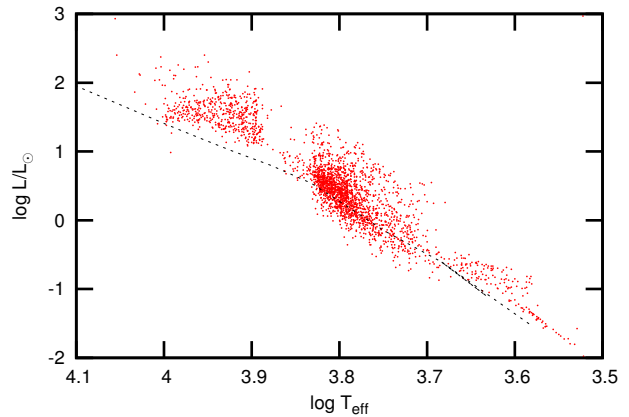}
\caption{The H-R diagram for stars with measured values of the differential
rotation shear.  The dashed line is the solar abundance zero-age main sequence
from \citet{Bertelli2008}.}
\label{hrdiag}
\end{figure}

\section{Normalization}

The frequency spread, $\delta\omega$ estimated from the time-frequency plots 
provides only a lower bound for the shear, $\Delta\Omega$. For two spots at
latitudes $\theta_1$ and $\theta_2$,
$\delta\omega = \Delta\Omega(|\sin^2\theta_2 - \sin^2\theta_1|) 
= f\Delta\Omega$.  The unknown factor $f$ depends on the spot locations.
\citet{Hall1994} made a simple estimate of $f$ on the assumption that
spots are equally spaced in latitude, in which case $f$ is easily calculated
as a function of the number of spots.  It turns out that $0.5 < f < 1.0$. 

In general, we do not know the number of spots or their distribution in
latitude unless the DI method is used.  Usually, discussions in the
literature are confined to the power laws describing how $\Delta\Omega$ varies 
with effective temperature or rotation rate, which do not require knowledge
of the scaling factor, $f$.  An alternative approach is to assume that
$\Delta\Omega$ in stars with approximately the same effective temperature and 
rotation period as the Sun will be the solar value, $\Delta\Omega_\odot = 
0.073$\,rad\,d$^{-1}$.  The normalization procedure consists in finding the 
mean of the frequency spread, $\langle\delta\omega\rangle$, for a group of 
stars with similar effective temperatures and rotation periods as the Sun and 
using $f = \langle\delta\omega\rangle/\Delta\Omega_\odot$.

To obtain the normalizing factor, we selected stars with $5300 < T_{\rm eff} 
< 6300$\,K  and $20 < P_{\rm rot} < 30$\,d from which we obtain
$\langle\delta\omega\rangle = 0.104 \pm 0.005$\,rad\,d$^{-1}$ and 
$f = 1.43 \pm 0.06$ from 82 stars.  The fact that $f > 1$ is a surprise.  
Perhaps the assumption that rotational shear in these stars is similar to
that in the Sun is not correct.  

\section{Results}

In Fig.\,\ref{hrdiag} we show the location of the stars in the H-R diagram.  
Effective temperatures, surface gravities and metallicities were taken from 
\citet{Huber2014}.  The luminosities are calculated using the relationships 
in \citet{Torres2010a}. The pulsating $\gamma$~Dor stars have multiple 
frequencies within the range of the expected rotational frequencies.  It is 
not possible to distinguish between rotational modulation and pulsation in 
these stars, so they were excluded.  A large fraction of F stars appear to be 
$\gamma$~Dor variables, which accounts for the gap in the 6500--7500\,K range 
in Fig.\,\ref{hrdiag}.

The A stars are usually omitted in discussions of starspots.  This
follows from the long-held view that radiative atmospheres cannot support a
magnetic field; hence no starspots should exist.  However, even a quick
inspection of the {\it Kepler} A-star light curves shows that this view is
not correct, as discussed in \citet{Balona2013c}.  In fact, the
time-frequency diagrams for A stars do not differ in any general way from
those of F, G, or K stars.  Fig.\,\ref{gray} includes the time-frequency
diagrams for two early A stars.

The properties of the sample of stars used in this investigation are shown in
Table\,\ref{stats}.  The number of stars with a measurable frequency spread
is given by $N$.  In some cases no frequency spread could be measured,
and $\Delta\Omega = 0$.  The number of these stars is given by $N_0$.  The 
mean values of the normalized shear and $\alpha$ are also shown.

\begin{table}
\begin{center}
\caption{The number of stars, $N$, within the given effective temperature 
range, and with $\Delta\Omega > 0$ is given.  $N_0$ is the number of stars 
with $\Delta\Omega = 0$.  The mean rotation period, $P_{\rm rot}$, the mean
rotational shear, $\Delta\Omega$, and the mean relative shear, $\alpha$ are
shown.  Values of $\Delta\Omega$ have been normalized to the Sun.}
\label{stats}
\begin{tabular}{lrrrrrr}
\hline
\multicolumn{1}{c}{Sp.Ty.} &
\multicolumn{1}{c}{$T_{\rm eff}$} &
\multicolumn{1}{c}{$N$} &
\multicolumn{1}{c}{$N_0$} &
\multicolumn{1}{c}{$P_{\rm rot}$} &
\multicolumn{1}{c}{$\Delta\Omega$} &
\multicolumn{1}{c}{$\alpha$} \\
\multicolumn{1}{c}{} &
\multicolumn{1}{c}{K} &
\multicolumn{1}{c}{} &
\multicolumn{1}{c}{} &
\multicolumn{1}{c}{d} &
\multicolumn{1}{c}{rad d$^{-1}$} &
\multicolumn{1}{c}{} \\
\hline
K9--G9 & 3600--5000  &  166 &   4 & 11.90 & 0.039 & 0.072  \\
K3--G4 & 4500--5600  &  272 &   4 & 12.09 & 0.068 & 0.126  \\
G7--G1 & 5200--5900  &  385 &   8 & 12.42 & 0.083 & 0.157  \\ 
G4--F8 & 5600--6200  &  604 &  20 & 11.08 & 0.112 & 0.181  \\
G0--F6 & 6000--6600  &  969 &  82 &  8.27 & 0.159 & 0.172  \\
F0--A0 & 7400--10000 &  315 & 207 &  3.41 & 0.263 & 0.066  \\
\hline
\end{tabular}
\end{center}
\end{table}

\begin{figure}
\centering
\includegraphics[]{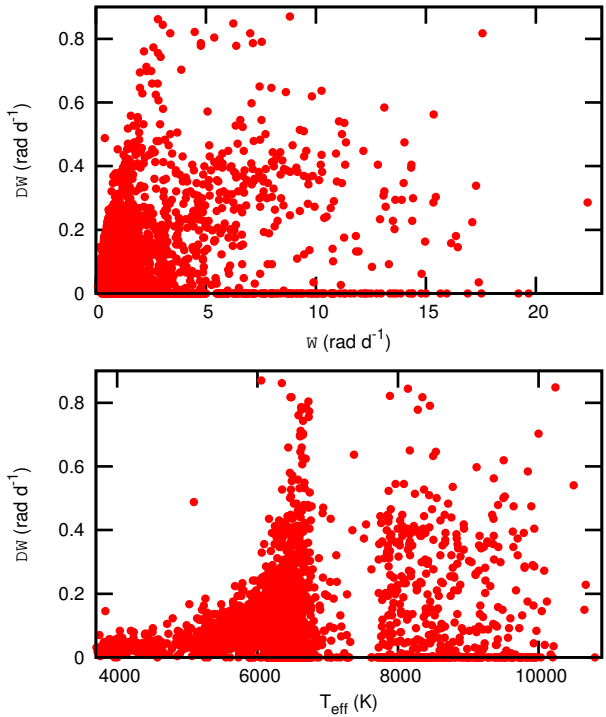}
\caption{Individual measurements of normalized differential rotation shear, 
$\Delta\Omega$, as a function of angular rotation rate (top panel)
and effective temperature (bottom panel).}
\label{temprot}
\end{figure}

\section{Variation of the shear with effective temperature and
rotation}

Fig.\,\ref{temprot} shows the normalized estimates of $\Delta\Omega$ 
as a function of effective temperature and rotation rate for each star.  
$\Delta\Omega$ increases with effective temperature up to the convective 
boundary, but decreases slightly for the A stars.  In general the shear
increases with increasing rotation rate.

It is evident that the shear is a complex function of both effective 
temperature and rotation rate. Since only a lower bound of $\Delta\Omega$ is 
measured, a relatively large number of stars within a given range of 
effective temperature and rotation rate is required to minimize
uncertainties.  Because of the small number of stars, previous attempts have 
used stars of all rotation rates to analyse the dependence of the shear on 
effective temperature.  In the same way, stars of all effective temperatures 
have been used to determine the dependence of shear on the rotation rate.  
This has prevented any meaningful estimate on how $\Delta\Omega$ varies 
independently with effective temperature and rotation rate.  The sample of
2562 {\it Kepler} stars is sufficiently large to solve this problem. 

\begin{figure}
\centering
\includegraphics[]{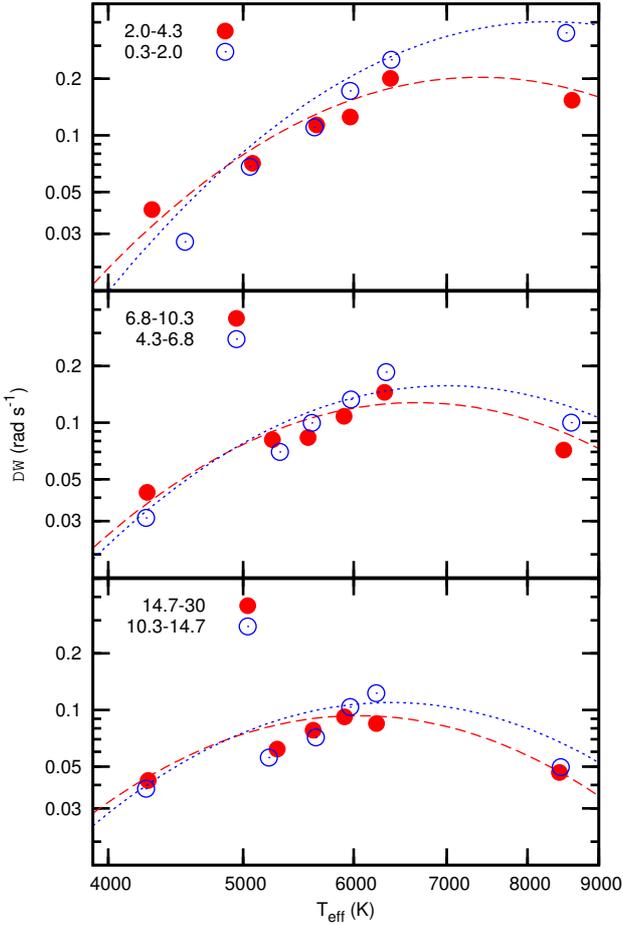}
\caption{The normalized differential rotation shear, $\Delta\Omega$, as a 
function of effective temperature for the indicated ranges of rotation period 
(in days). The curves are calculated using the interpolation formula
Eq.\,\ref{Eq1}.}
\label{tempdomg}
\end{figure}

The relationship between the shear and effective temperature needs to be 
evaluated for constant values of rotation rate, $\Omega$.  For this purpose
the mean rotation frequency spread of stars within a small range of 
$\Omega$ was calculated.   Stars with zero frequency spread were not
included.  Fig.\,\ref{tempdomg} shows how $\Delta\Omega$ in stars with
approximately the same rotational period varies with effective temperature.

\begin{figure}
\centering
\includegraphics[]{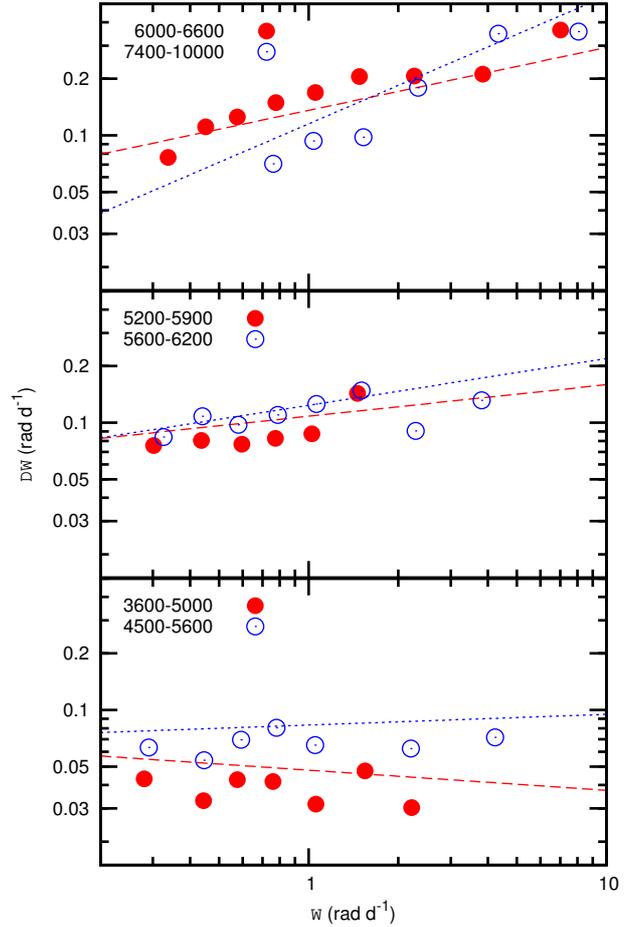}
\caption{The variation of $\Delta\Omega$ as a function of angular rotation 
frequency for the indicated ranges of effective temperature. The solid 
curves are calculated using the interpolation formula.}
\label{rotfdomg}
\end{figure}

In the same way, to evaluate the relationship between the shear and rotation 
rate, the mean rotation frequency spread of stars within a small range of 
effective temperature was calculated.  Fig.\,\ref{rotfdomg} shows how
$\Delta\Omega$ in stars of about the same effective temperature varies with
rotation rate.

The figures show that $\Delta\Omega$ is weakly dependent on effective
temperature for slowly rotating stars, but the dependence increases
with increasing rotation rate.  For K and G stars there is very little
dependence of $\Delta\Omega$ on rotation rate, but this dependence increases
for F and A stars.  The fact that the shear depends on effective temperature 
and rotation rate in a rather complex way probably explains why there is so 
much disagreement in the literature. 

A reasonable fit to the data may be obtained using the following
interpolation formula:
\begin{align}
\log\Delta\Omega &= -211.12 + 109.67x - 11.14y -14.30x^2 + 3.02xy,
\label{Eq1}
\end{align}
where $x = \log T_{\rm eff}$ and $y = \log\Omega$ with $\Omega$ and
$\Delta\Omega$ in rad\,d$^{-1}$.  In Figs.\,\ref{tempdomg} and
\ref{rotfdomg}, the curves show $\Delta\Omega$ calculated from this formula.

\section{Comparison with previous observations}

Several observational studies have found a power law of the form
$\Delta\Omega \propto \Omega^n$.  By measuring the change in rotation 
periods in 36 F--K stars, \citet{Donahue1996} found that $n \approx 0.7$.
\citet{Messina2003} estimated $n \approx 0.6$ from the period variation in
14 K5--G2 stars.   \citet{Barnes2005} find $n \approx 0.15$ by 
combining their DI results of 10 stars with previous studies.  Using the FT 
method, \citet{Reiners2006} find $n \approx 1.5$ from 10 F and G stars.
Using the interpolation formula we get $n = -0.1$, 0.2, 0.4 and 0.7 for K,
G, F and A stars respectively.  It is evident that the relationship depends 
rather sensitively on the effective temperature.

There is also a power law relating $\Delta\Omega$ to effective temperature
of the form $\Delta\Omega \propto T_{\rm eff}^p$.  \citet{Barnes2005} found 
$p \approx 8.9$.  The interpolation formula gives $p = 6.4$, 3.5, 1.0 and
-2.0 for K, G, F and A stars.  It is clearly necessary to disentangle the
temperature and rotation rate variation for a proper description of the
shear.

\begin{table}
\begin{center}
\caption{Measurements of rotational shear, $\Delta\Omega$, obtained by
Doppler imaging.  The star name and effective temperature are given.  
References are as follows: 
 1 - \citet{Marsden2004};  2 - \citet{Marsden2005};  3 - \citet{Donati2000}; 
 4 - \citet{Barnes2000};   5 - \citet{Collier2002b}; 6 - \citet{Donati2003}; 
 7 - \citet{Barnes2005};   8 - \citet{Barnes2005b};  9 - \citet{Barnes2004}; 
10 - \citet{Jeffers2009}; 11 - \citet{Jeffers2008}; 12 - \citet{Jarvinen2015}; 13 - \citet{Kovari2014}; 14 - \citet{Fares2012};
15 - \citet{Kovari2011};  16 - \citet{Waite2011};   17 - \citet{Marsden2011a};
18 - \citet{Marsden2011b}; 19 - \citet{Dunstone2008}.}
\label{domg}
\begin{tabular}{lrlrr}
\hline
\multicolumn{1}{c}{Star} &
\multicolumn{1}{c}{$T_{\rm eff}$} &
\multicolumn{1}{c}{$\Delta\Omega$} &
\multicolumn{1}{c}{$P_{\rm rot}$} &
\multicolumn{1}{c}{Ref.} \\
\multicolumn{1}{c}{} &
\multicolumn{1}{c}{K} &
\multicolumn{1}{c}{rad d$^{-1}$} &
\multicolumn{1}{c}{d} \\
\hline
HD 307938  &  5859 &  $0.025 \pm 0.015$ &  0.57 &  1 \\
HD 307938  &  5859 &  $0.140 \pm 0.010$ &  0.57 &  2 \\ 
LQ Lup     &  5729 &  $0.130 \pm 0.020$ &  0.31 &  3 \\ 
PZ Tel     &  5448 &  $0.101 \pm 0.007$ &  0.95 &  4 \\ 
AB Dor     &  5386 &  $0.046 \pm 0.006$ &  0.51 &  5 \\
AB Dor     &  5386 &  $0.091 \pm 0.012$ &  0.51 &  5 \\      
AB Dor     &  5386 &  $0.089 \pm 0.008$ &  0.51 &  5 \\ 
AB Dor     &  5386 &  $0.067 \pm 0.020$ &  0.51 &  5 \\  
AB Dor     &  5386 &  $0.071 \pm 0.006$ &  0.51 &  5 \\ 
AB Dor     &  5386 &  $0.058 \pm 0.005$ &  0.51 &  5 \\
AB Dor     &  5386 &  $0.053 \pm 0.003$ &  0.51 &  6 \\ 
AB Dor     &  5386 &  $0.047 \pm 0.003$ &  0.51 &  6 \\
AB Dor     &  5386 &  $0.058 \pm 0.002$ &  0.51 &  6 \\
AB Dor     &  5386 &  $0.046 \pm 0.003$ &  0.51 &  6 \\ 
AB Dor     &  5386 &  $0.054 \pm 0.001$ &  0.51 &  6 \\ 
HD 197890  &  4989 &  $0.032 \pm 0.002$ &  0.38 &  7 \\ 
LQ Hya     &  5019 &  $0.194 \pm 0.022$ &  1.60 &  6 \\ 
LQ Hya     &  5019 &  $0.014 \pm 0.003$ &  1.60 &  6 \\
LO Peg     &  4577 &  $0.036 \pm 0.007$ &  0.42 &  8 \\
HK Aqr     &  3697 &  $0.005 \pm 0.009$ &  0.43 &  9 \\ 
EY Dra     &  3489 &  $0.000 \pm 0.003$ &  0.46 &  7 \\
HD 171488  &  5800 &  $0.340 \pm 0.040$ &  1.33 & 10 \\
HD 171488  &  5800 &  $0.402 \pm 0.044$ &  1.33 & 10 \\
HD 171488  &  5800 &  $0.470 \pm 0.044$ &  1.33 & 11 \\
AF Lep     &  6100 &  $0.259 \pm 0.019$ &  0.97 & 12 \\  
IL HYa     &  4500 &  $0.035 \pm 0.003$ & 12.73 & 13 \\ 
HD 179949  &  6160 &  $0.216 \pm 0.061$ &  7.62 & 14 \\
V889 Her   &  5750 &  $0.042$           &  1.34 & 15 \\
HD 106506  &  5900 &  $0.240 \pm 0.030$ &  1.39 & 16 \\
HD 106506  &  5900 &  $0.210 \pm 0.030$ &  1.39 & 16 \\
HD 141943  &  5850 &  $0.240 \pm 0.030$ &  2.17 & 17 \\
HD 141943  &  5850 &  $0.360 \pm 0.090$ &  2.17 & 18 \\
HD 141943  &  5850 &  $0.450 \pm 0.080$ &  2.17 & 18 \\
HD 155555A &  5400 &  $0.078 \pm 0.006$ &  1.67 & 19 \\
HD 155555A &  5400 &  $0.143 \pm 0.008$ &  1.67 & 19 \\
HD 155555A &  5400 &  $0.104$           &  1.67 & 19 \\
HD 155555B &  5050 &  $0.039 \pm 0.006$ &  1.67 & 19 \\
HD 155555B &  5050 &  $0.088 \pm 0.006$ &  1.67 & 19 \\
HD 155555B &  5050 &  $0.060$           &  1.67 & 19 \\
\hline
\end{tabular}
\end{center}
\end{table}

The DI technique provides a direct method to estimate $\Delta\Omega$.  
Table\,\ref{domg}, which is an update of Table\,1 of \citet{Barnes2005},
lists stars for which $\Delta\Omega$ has been determined.  It is interesting
to compare these values with the values predicted from Eq.\,\ref{Eq1}.
It turns out that the average difference in shear between the stars observed
by the DI method and that calculated from the interpolating formula is
$\langle\Delta\Omega_{\rm DI} - \Delta\Omega_{\rm Eq1}\rangle = 0.000 
\pm 0.018$\,rad\,d$^{-1}$.  In other words, there is no significant 
difference between the two values of $\Delta\Omega$.  Also, there is no 
discernible systematic trend in this difference as a function of effective 
temperature or rotation period.  This suggests that the adopted normalizing
factor, $f$, is probably correct.  It also suggests that differential rotation
in these heavily-spotted stars is similar to that in the Sun.

\begin{figure}
\centering
\includegraphics[]{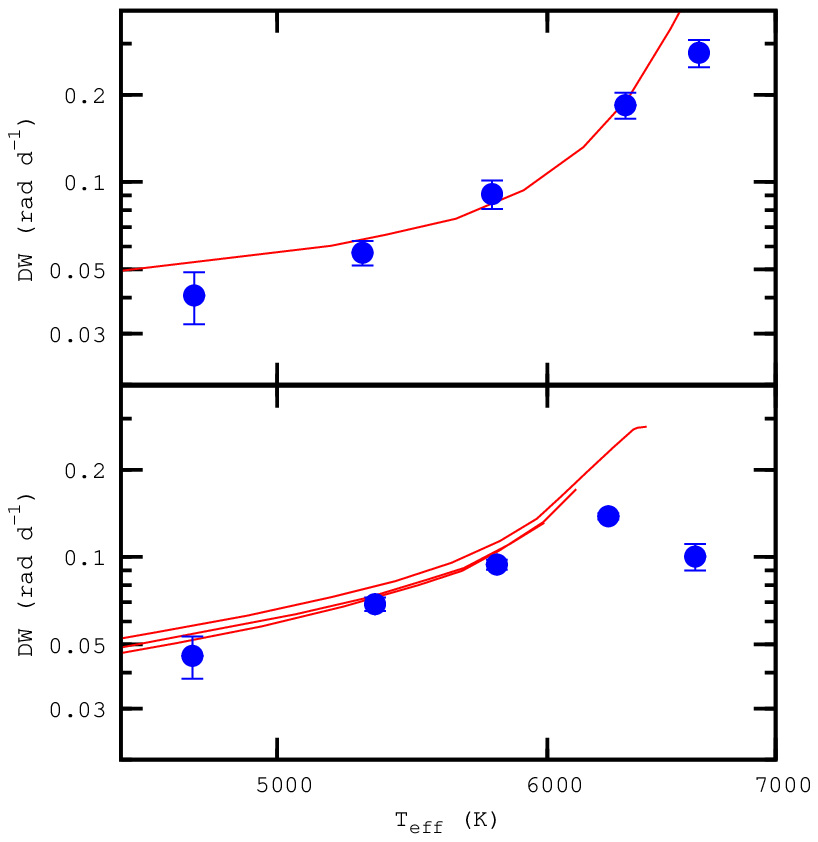}
\caption{Top panel: the solid line is the variation of $\Delta\Omega$ with 
effective temperature for models with a rotation period of 
$P_{\rm rot} = 2.5$\,d by \citet{Kuker2011}. The points are the observed 
values for stars with $2.0 < P_{\rm rot} < 3.0$\,d.  Bottom panel: The lines 
are models with abundances of $Z = 0.01, 0.02$ and 0.03 and a rotation period 
of 10\,d by \citet{Kitchatinov2012}.  The points are the observed values with 
$6.0 < P_{\rm rot} < 14.0$\,d. Error bars are one standard deviation in 
length.}
\label{kukerteff}
\end{figure}

\begin{figure}
\centering
\includegraphics[]{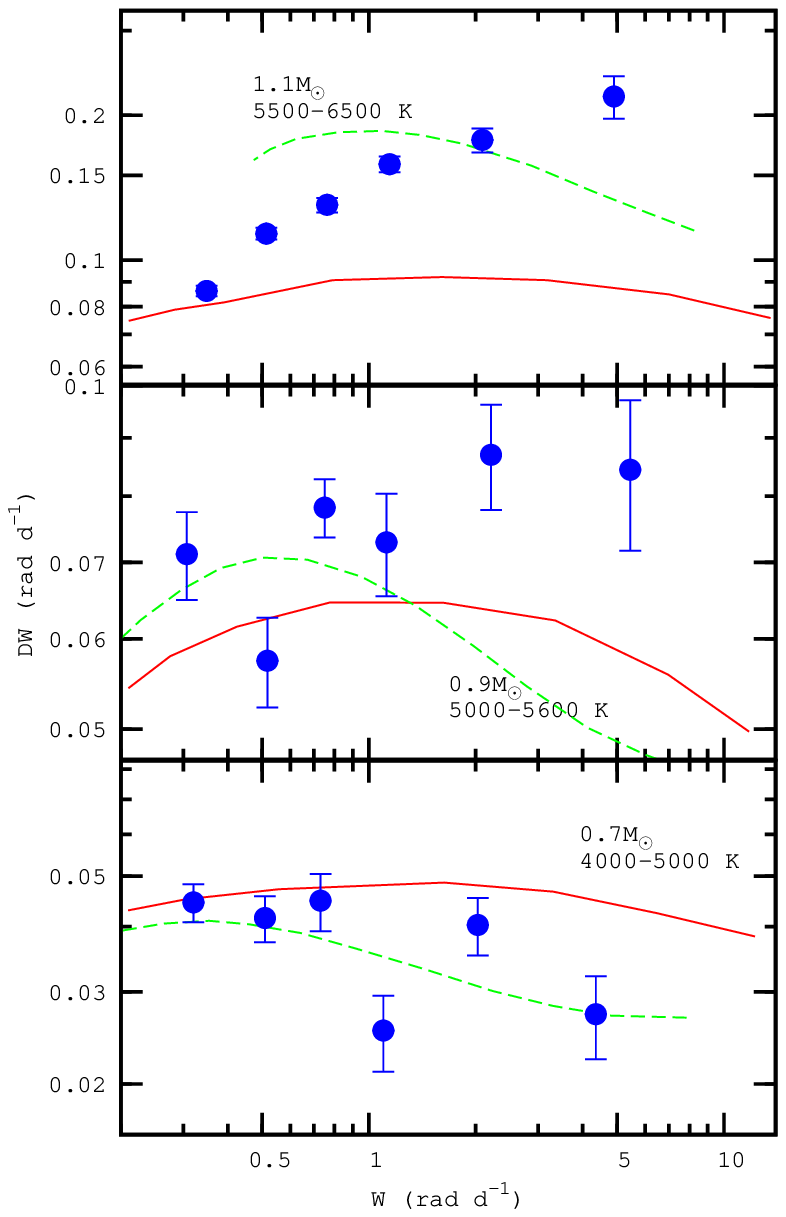}
\caption{The solid lines are the variation of $\Delta\Omega$ with angular
rotation rate for models of stars by \citet{Kuker2011} with the indicated 
masses. The dashed lines are the corresponding models by 
\citet{Kitchatinov2012}.  The filled circles are normalized values
of $\Delta\Omega$ for stars in the indicated range of effective
temperature. Error bars are one standard deviation in length.}
\label{kukerprot}
\end{figure}

\section{Comparison with theory}

Rotation reduces the effective gravity at the equator which, in turn, leads
to a temperature variation from equator to pole.  This creates a thermal 
imbalance which is the cause of meridional circulation.  Angular momentum 
transport by convection and by the meridional flow together with the
Coriolis force produces differential rotation. Differential rotation of 
main-sequence dwarfs is predicted to vary mildly with rotation rate but to 
increase strongly with effective temperature.  Meridional circulation and 
differential rotation are key ingredients in the current theory of the solar 
dynamo \citep{Choudhuri1995, Kuker2001}. 

Models of stellar differential rotation for F, G, K and M dwarfs were
computed by \citet{Kuker2005} by solving the equation of motion and the 
equation of convective heat transport in a mean-field formulation. 
For each spectral type, the rotation rate is varied to study the dependence of 
the surface shear on this parameter.  The horizontal shear, $\Delta\Omega$, 
turns out to depend strongly on the effective temperature and only weakly on 
the rotation rate. Later, \citet{Kuker2007} calculated differential rotation
specifically for F stars.  These models show signs of very strong differential
rotation in some cases. Stars just cooler than the granulation boundary
have shallow convection zones with short convective turnover times.  This
leads to a horizontal shear that is much larger than on the solar surface, 
in agreement with observations.

More recently, \citet{Kuker2011} have further explored the variation of 
surface differential rotation and meridional flow along the lower part of the 
zero age main sequence.  They construct mean field models of the outer 
convection zones and compute differential rotation and meridional flow by
solving the Reynolds equation with transport coefficients from the second
order correlation approximation.  For a fixed rotation period of 2.5\,d,
they find a strong dependence of $\Delta\Omega$ on the effective
temperature, which is weak in M dwarfs and rises sharply for F stars
(top panel of Fig.\,\ref{kukerteff}). The increase with effective temperature 
is modest below  6000\,K but very steep above 6000\,K.  Both the surface 
rotation and the meridional circulation are solar-type over the entire 
temperature range.  

They also study the dependence of $\Delta\Omega$ on the rotation rate.  This
dependence is weak (Fig.\,\ref{kukerprot}).  Numerical experiments show that 
for effective temperatures below 6000\,K the Reynolds stress is the dominant 
driver of differential rotation.  At this time, there has been no numerical 
study of differential rotation for A stars.

In Fig.\,\ref{kukerteff} we show how $\Delta\Omega$ varies with $T_{\rm eff}$ 
for stars with rotation periods in the range 2.0--3.0\,d. There 
is good agreement with the models of \citet{Kuker2011}.  Fig.\,\ref{kukerprot} 
shows how $\Delta\Omega$ varies with rotation rate for {\it Kepler} stars 
within limited effective temperature ranges.  There agreement with the
\citet{Kuker2011} models is reasonable for K stars but there are
significant departures for G stars.  For F stars the observations depart
quite strongly from the models, particularly at high rotation rates.  In the
models $\Delta\Omega$ attains a maximum at about $\Omega \approx
2$\,rad\,d$^{-1}$, whereas the observations show that it increases
monotonically with increasing rotation rate. 

\citet{Kitchatinov2012} have also calculated models of differentially
rotating stars using the mean field formulation.  They find that the
dependence of $\Delta\Omega$ on metallicity for stars of a given mass is 
quite pronounced. However, the dependence almost disappears when differential 
rotation is considered as a function of effective temperature.  Their models 
of $\Delta\Omega$ as a function of effective temperature for a fixed rotation 
period of 10\,d is compared with observations in the bottom panel of 
Fig.\,\ref{kukerteff}.  The agreement is good except for the very hottest
stars.  \citet{Kitchatinov2012} also computed models of how $\Delta\Omega$ 
varies with rotation rate for models with different masses and metallicity 
$Z = 0.02$.  The results are shown in Fig.\,\ref{kukerprot}.  As with
\citet{Kuker2011}, there is poor agreement with observations except for the
K stars.

\citet{Augustson2012} use a three-dimensional anelastic spherical harmonic 
code to simulate global-scale turbulent flows in 1.2 and $1.3M_\odot$ F-type 
stars at varying rotation rates.  They find that differential rotation becomes
much stronger with more rapid rotation and larger mass, such that the
rotational shear between the equator and a latitude of $60^\circ$ is
$\Delta\Omega_{60} = 0.083 \left(\frac{M}{M_\odot}\right)^{3.9}
\left(\frac{\Omega}{\Omega_\odot}\right)^{0.6}$\,rad\,d$^{-1}$.  The model
roughly corresponds to a main sequence star with $T_{\rm eff} \approx
6500$\,K.

\begin{figure}
\centering
\includegraphics[]{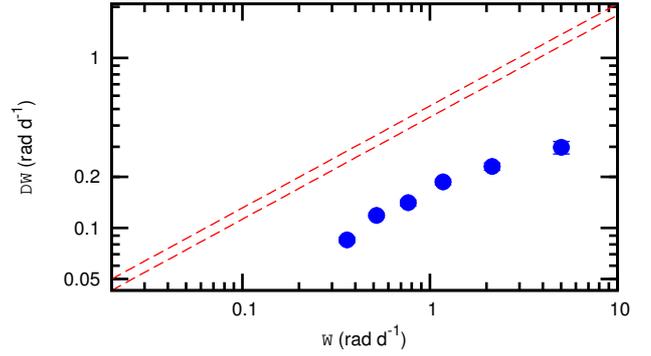}
\caption{The dashed lines are the $\Delta\Omega_{60}$ relationships of 
\citet{Augustson2012} for $M/M_\odot = 1.2$ (bottom line) and 1.3 (top
line).  The filled circles are the observed values for $6000 < T_{\rm eff} 
< 7000$.}
\label{augustson}
\end{figure}

In Fig.\,\ref{augustson} we show a comparison of this relationship with
observations.  The slope is in good agreement, though the constant factor is 
too large for the models.

\begin{figure}
\centering
\includegraphics[]{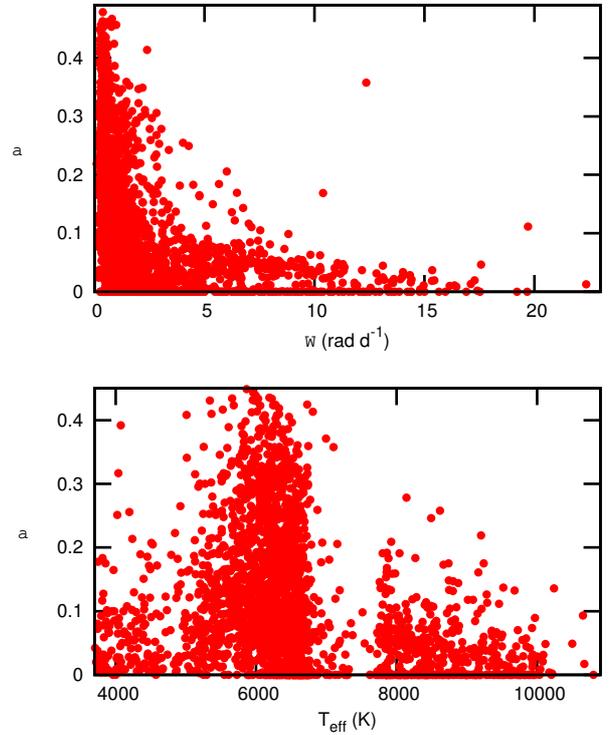}
\caption{Individual measurements of the relative differential rotation shear, 
$\alpha$, as a function of effective temperature (bottom panel) and angular 
rotational rate (top panel).}
\label{rattemp}
\end{figure}

In summary, it appears that all current models describe the variation of
$\Delta\Omega$ with effective temperature rather well.  Although models
using the mean field formulation seem to describe the variation of 
$\Delta\Omega$ with rotation rate quite well for K stars, problems begin with 
the G stars and are very severe in the F stars.  The models of 
\citet{Augustson2012} for F stars gives the correct power law for
$\Delta\Omega$ as a function of rotation rate, but the shear is consistently
too high. 

\section{The relative shear, $\alpha$}

The relative shear is given by $\alpha = \Delta\Omega/\Omega_e$.  For each 
star in our sample we use the normalized value of $\Delta\Omega$ and the
known rotation period to obtain $\alpha$.  Fig.\,\ref{rattemp} shows
individual values of $\alpha$ as a function of effective temperature and 
rotation rate.

\begin{figure}
\centering
\includegraphics[]{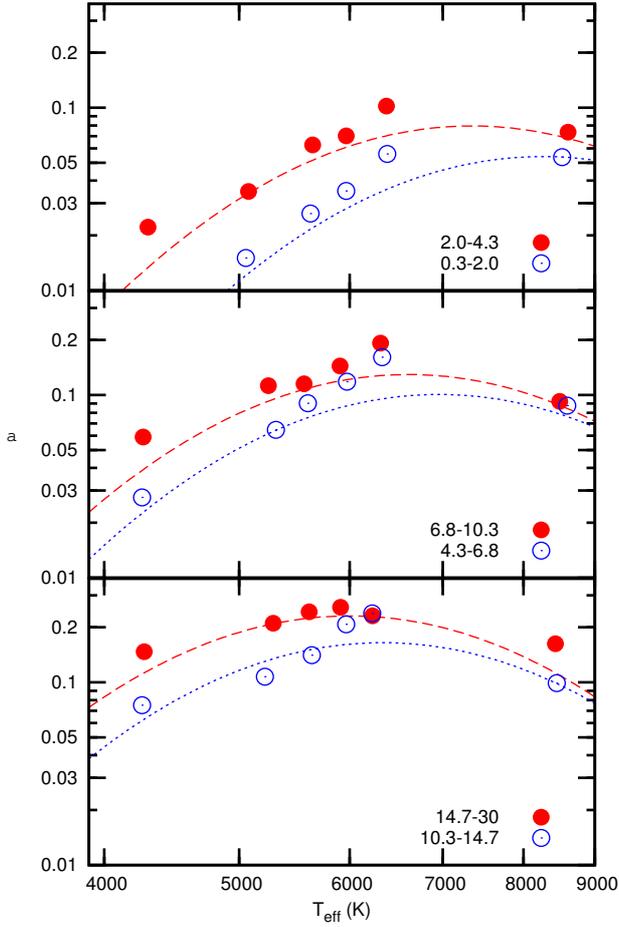}
\caption{The normalized relative rotational shear, $\alpha$, as a 
function of effective temperature for the indicated ranges of rotation period 
(in days). The curves are calculated using the interpolation formula.}
\label{tempdrat}
\end{figure}

\begin{figure}
\centering
\includegraphics[]{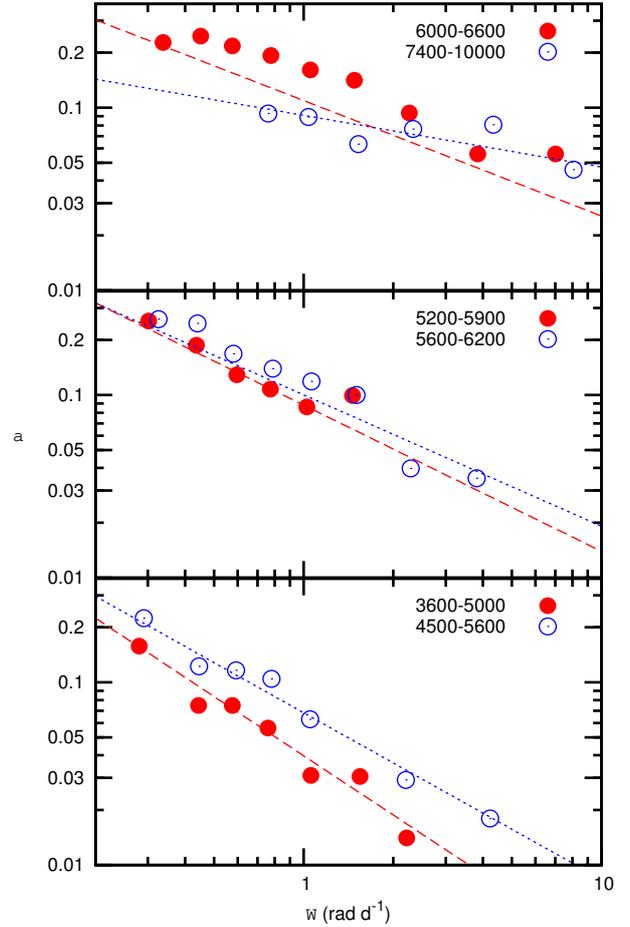}
\caption{The variation of $\alpha$ as a function of angular rotation 
frequency for the indicated ranges of effective temperature. The curves are 
calculated using the interpolation formula.}
\label{rotfdrat}
\end{figure}

As before, the mean value of $\alpha$ within narrow ranges of rotation 
period are calculated as a function of effective temperature.  
Fig.\,\ref{tempdrat} shows the results.  In the same way, the mean value of 
$\alpha$ within  narrow temperature ranges are calculated as a function of
rotation rate.  This is shown in Fig.\,\ref{rotfdrat}.

A suitable interpolation formula is given by: 
\begin{align}
\log\alpha &= -210.91 + 109.59x - 12.22y -14.30x^2 + 3.05xy,
\label{Eq2}
\end{align}
where $x = \log T_{\rm eff}$ and $y = \log\Omega$ with $\Omega$ and
$\Delta\Omega$ in rad\,d$^{-1}$.  The fit to the observations are shown in
Figs.\,\ref{tempdrat} and \ref{rotfdrat} by the lines.

Notice that the maximum value of $\alpha$ is less than 0.3 for all stars and 
decreases quite sharply with increasing rotation rate.  This is in contrast to
the absolute shear which stays the same or increases with rotation rate.  The 
reason for the different behaviour is the rapid increase of rotation rate,
$\Omega_e$, with effective temperature.  For example, the typical rotation
period of A stars is about 3\,d, while it is around 10\,d for K and G stars.

\section{The A stars}

Because of the perception that starspots cannot exist on stars with radiative 
envelopes, the A and B stars have been omitted from differential rotation
studies using the DI and photometric techniques.  The FT method, which 
measures the shape of the line profile, does not require the presence of
starspots.  Using this method, \citet{Reiners2004} and \citet{Reiners2006}
were able to detect differential rotation in three mid- to late-A stars.  
Using the same method, \citet{AmmlervonEiff2012} confirmed differential 
rotation among the same three A stars near the granulation boundary, but 
not in other A stars.  In fact, no differential rotation was found in stars 
with $T_{\rm eff} > 7400$\,K.  This is in contrast to our results which 
clearly show differential rotation in more than half of the {\it Kepler} A 
stars.  

It should be noted that the FT method measures $\alpha$, the dimensionless 
differential rotation constant and not the absolute shear, $\Delta\Omega$.  
We have seen that the value of $\alpha$ is small for the A stars; in fact 
$\alpha < 0.1$ for nearly all A stars.  \citet{AmmlervonEiff2012} state that
the detection limit for $\alpha$ is $\approx 0.05$.  This means that  most of 
the A stars that were observed by them are probably close to the detection 
limit for differential rotation.  This may be the reason for their failure to 
detect differential rotation for stars hotter than 7400\,K.

\begin{figure}
\centering
\includegraphics[]{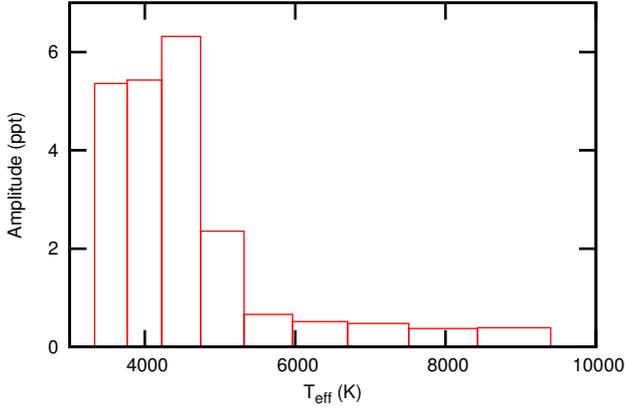}
\caption{The distribution of rotational modulation amplitude (in parts per
thousand) as a function of effective temperature.}
\label{ampdist}
\end{figure}

In Fig.\,\ref{ampdist} the average rotational light modulation amplitudes 
are shown as a function of effective temperature.  The amplitudes are large
in the K and G stars, but fall sharply for the F and A stars.  Let us
suppose that the temperature difference between the spot and surrounding
photosphere, $\Delta T$, is a constant for stars of all spectral types.  It
then follows that the light amplitude, which is approximately proportional to
$(\Delta T/T)^4$, must decrease with increasing temperature, $T$.  This
contrast effect is most likely responsible for the light amplitude decrease 
from G to A stars.  The rotational light amplitudes for A stars are less than 
1\,millimag and not detectable from ground-based observations.  This is
possibly why rotational modulation in A stars escaped detection until 
recently.

\begin{figure}
\centering
\includegraphics[]{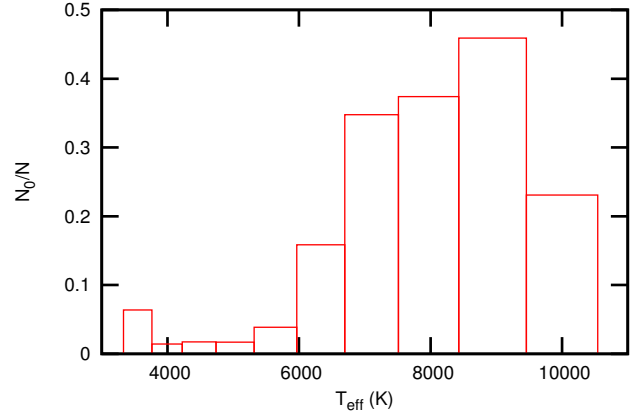}
\caption{The fraction of stars with $\Delta\Omega = 0$ as a function of 
effective temperature.}
\label{zerdist}
\end{figure}

The large fraction of A stars with no detectable rotational frequency spread, 
i.e. with $\Delta\Omega = 0$ (see Table\,\ref{stats}) deserves attention.   In
Fig.\,\ref{zerdist} the relative number of stars with $\Delta\Omega = 0$ is
shown as a function of effective temperature.  It is evident that the numbers 
of A stars with zero frequency spread is large.  In fact, about 40\,per\,cent 
of A stars showing rotational modulation do not have measurable frequency 
spread.  This fraction decreases sharply for the F stars (about
25\,per\,cent) and is less than 5\,per\,cent for cooler stars.

There is no tendency for stars with zero frequency spread to have significantly
different light modulation amplitudes or different rotation rates.  There
appears to be two distinct groups among the A and F stars: the majority with 
large differential rotation shear and a smaller group with no shear at all
or spots confined to a narrow range of latitude.  It would be important to
use the FT method with more precision to decide which of the two
explanations is correct.

\section{Summary and discussion}

We used {\it Kepler} photometry of stars with rotational modulation to
measure the frequency spread, $\delta\omega$, around the rotational period.  
The frequency spread, which is assumed to be a result of starspots at 
different latitudes, provides a lower bound to the absolute rotational
shear, $\Delta\Omega$.  Estimation of frequency spread was made by visual 
examination of time-frequency grayscale diagrams.  We were careful to avoid 
pulsating stars, eclipsing binaries or other types of variable which might be 
mistaken for rotational modulation.  We also confined the sample to stars with
known rotational periods.  In this way we obtained estimates of $\delta\omega$
for 2562 {\it Kepler} stars, including A stars.

The frequency spread is related to the rotational shear by $\delta\omega = 
f\Delta\Omega$.  If it is assumed that the distribution of spots in latitude 
is approximately the same in all stars, then the average value of 
$f$ is independent of surface temperature or rotation period.  Since
$\delta\omega$ provides a lower bound to $\Delta\Omega$, we expect $f < 1$.
Furthermore, if we assume that surface differential rotation in the Sun is 
typical of stars with about the same effective temperature and rotation
period, then we can obtain $f$ from the mean value of $\delta\omega$ of
these stars and the solar value of $\Delta\Omega$.  It is found that 
$f = 1.43$, which is puzzling because the frequency spread should be lower 
than the rotational shear.  We suspect that one of the assumptions is not 
correct.  It is possible that $\Delta\Omega$ may be systematically larger in
these stars which have spots that are at least 100 times larger than in the 
Sun.

To investigate the functional dependence of $\Delta\Omega$ on effective
temperature we determined $\Delta\Omega$ in restricted ranges of rotation
period.  In like manner, we found the functional dependence of the shear on 
the rotation period by determining $\Delta\Omega$ in restricted ranges of 
effective temperature.  The result can be expressed by the interpolation
formula, Eq.\,\ref{Eq1}.  It is possible that not all stars obey this
relationship exactly, since $\Delta\Omega$ probably depends on the history
of the star.  Eq.\,\ref{Eq1} should be taken as an approximation which
applies, on average, to a group of stars with similar effective temperatures 
and rotation periods.

The discrepancies which arise in the literature regarding the power law
dependence of $\Delta\Omega$ on effective temperature and rotation rate are
most likely a result of an insufficient number of stars with known values of
rotational shear.  We compared $\Delta\Omega$ obtained with the Doppler imaging 
technique with Eq.\,\ref{Eq1}.  There is no significant difference between 
the two values of $\Delta\Omega$, which provides confidence in the procedure
of using the solar value to obtain the normalizing factor, $f$.

We compared the observed values of $\Delta\Omega$ with those predicted by
the models of \citet{Kuker2011, Kitchatinov2012} and \citet{Augustson2012}.
The models provide a good description of how $\Delta\Omega$ varies with 
effective temperature.  However, the models predict a variation with rotation 
rate which differs from the observed values for G and F stars.  Whereas the 
models predict a decrease in $\Delta\Omega$ for rotation periods shorter than 
about 6\,d, the observations show that the shear keeps on increasing towards 
shorter periods for G and F stars.  The discrepancy is very large for F
stars except in the models by \citet{Augustson2012} which predicts the
correct power law.

We also studied the relative shear, $\alpha$, which reaches a maximum for F
stars.  For A stars $\alpha < 0.1$, a value which is near the detection limit 
of the Fourier transform method.  This may explain the discrepancy in the
estimation of $\Delta\Omega$ for A stars between \citet{AmmlervonEiff2012}
and the results presented here.  We provide an interpolation formula
(Eq.\,\ref{Eq2}) which allows $\alpha$ to be calculated given the effective
temperature and rotation rate.

The number of stars with no detectable rotational shear is largest among the
A stars, comprising nearly 40\,per\,cent of the sample.  This number is
about 25\,per\,cent for F stars but drops to less than 5\,per\,cent for G and 
K stars.  Perhaps there are two populations of F and A stars: a minority with 
rigid body rotation and a majority with large rotational shear.  Alternatively,
starspots may be restricted to a relatively small range of latitude in some
F and A stars.  Observations using the Fourier transform method applied with 
greater precision would be able to resolve this problem.

One of the most important aspects of this study is the fact that rotational
shear in A stars is clearly observed.  The time-frequency plots show
the same pattern of spot migration and spot lifetimes in A stars as in the
cooler stars.  Moreover, the functional behaviour of absolute and relative
shear in A stars is a smooth extension of that in K, G and F stars.  The fact 
that photometric rotation periods agree with those determined from 
spectroscopic measurements of projected rotational velocity 
\citep{Balona2013c}, shows that the variability in A stars is indeed a
result of rotational modulation.  Recently, \citet{Bohm2015} have detected 
starspots in the hot A star Vega using high-dispersion spectroscopy.  The 
long-held notion that spots should not exist in stars with radiative envelopes
is clearly not correct.  It is likely that starspots are also present in B 
stars.  The few {\it Kepler} observations of B stars show that rotational 
modulation is probably present in nearly half of these stars 
\citep{Balona2011b, Balona2015c, Balona2016a}. 

There are further indications that the current view of A star atmospheres is 
not correct.  Space photometry has shown that all $\delta$~Scuti stars
pulsate with both high and low frequencies \citep{Balona2015d}, but models are
unable to account for low-frequency pulsations.  Multiple low frequencies
are, however, a characteristic feature of the cooler $\gamma$~Doradus stars.
The pulsations in these F stars are driven by the convective blocking 
mechanism \citep{Guzik2000}.  In other words, pulsations in the A-type 
$\delta$~Scuti stars behave as if they have convective envelopes.  

A study of differential rotation in A stars does not yet exist.  Although
hardly surprising, this serious shortcoming may hopefully be addressed soon.  
In A stars, the shear is very sensitive to rotation rate, increasing rather 
steeply with increasing rotation rate. Such a study may shed light on the 
nature of A star atmospheres from the behaviour of $\Delta\Omega$ with 
effective temperature and rotation rate.

\section*{Acknowledgments}

LAB wishes to thank the National Research Foundation of South Africa for 
financial support.  OPA acknowledges funding from Material
Science Innovation and Modelling (MaSIM) Research Focus area - North West
University, South Africa.

\bibliographystyle{mn2e}
\bibliography{difrot}

\label{lastpage}

\end{document}